\documentclass[a4paper,aps,prl,twocolumn,showpacs]{revtex4}

\usepackage[dvips]{graphicx}
\usepackage{amsmath}
\usepackage{amssymb}
\usepackage{natbib}
\usepackage{psfrag}
\usepackage{wasysym}

\newcommand{\ta}{\sigma}
\newcommand{\T}{\Sigma}
\newcommand{\lgr}{\ell_g}
\newcommand{\upd}{\mathrm{d}}
\newcommand{\dl}{\mbox{$\Delta l$}}
\newcommand{\dL}{\mbox{$\Delta L$}}

\begin{document}

\title{Statics and Inertial Dynamics of a Ruck in a Rug}
\author{Dominic Vella$^{1,2}$,  Arezki Boudaoud$^1$ and Mokhtar Adda-Bedia$^1$}
\affiliation{$^1$ Laboratoire de Physique Statistique, Ecole Normale Sup\'{e}rieure, UPMC  
Paris 06, Universit\'{e} Paris Diderot, CNRS, 24 rue Lhomond, 75005 Paris,  
France \\
$^2$ ITG, Department of Applied Mathematics and Theoretical Physics, University of Cambridge, Wilberforce Road, Cambridge, CB3 0WA, U.~K. }

\date{\today}

\begin{abstract}
We consider the familiar problem of a bump, or ruck, in a rug. Under lateral compression, a rug bends out of the plane forming a ruck --- a localized region in which it is no longer in contact with the floor. We show that when the external force that created the ruck is removed, the ruck flattens out unless the initial compression is greater than a critical value, which we determine. We also study the inertial motion of a ruck that is generated when one end of the rug is moved rapidly. We show that the equations of motion admit a travelling ruck solution for which a linear combination of the tension and kinetic energy is determined by the ruck size. We confirm these findings experimentally. We end by discussing the potential implications of our work for the analogous propagation of localized slip pulses in the sliding of two bodies in contact.
\end{abstract}



\maketitle

As well as being something of an annoyance in everyday life, the familiar bumps (or rucks) that form in a rug as people walk over it has long proved to be a useful analogy in explaining a range of important physical phenomena. For example, several authors have used the motion of a ruck to illustrate how dislocations can facilitate the relative motion of two crystalline planes of atoms (see pg.~117 of \cite{mott86}). In this situation the ruck is quasistatic. However, those working with carpets every day know that shaking one end leads to the propagation of rucks, which facilitate small-scale sliding \cite{comninou78}. This dynamic scenario is used to explain the observation of Schallamach waves~\cite{schallamach71}: waves of detachment that control the sliding of rubber surfaces~\cite{briggs75,barquins85}. In much the same way the slip pulses observed in some earthquakes~\cite{heaton90,adda03} are often thought to be akin to moving rucks in a rug~\cite{hough02}. 

Though these analogies are frequently referred to in the literature, we are not aware of any investigation of their quantitative validity nor even a study of the properties of rucks in rugs themselves. In this Letter, we consider the statics and inertial dynamics of a ruck in a rug. Our aim is to characterize the model system as a first step towards understanding the validity of these analogies. 

Recently, a great deal of attention has been focussed on understanding the statics and dynamics of thin elastic objects, of which a rug is a particular example. Static localized structures qualitatively similar to a ruck in a rug have been observed in the compression of thin films floating on water~\cite{luka08} or deadhering from a polymer substrate~\cite{vella09}. Generally, studies of the dynamics of thin elastic objects have been focused on investigating the interaction of a flexible object with a fluid \cite{hosoi04,argentina07}. The purely elastic propagation of localized disturbances along thin objects has received less attention, notable exceptions being investigations of a wave in a whip~\cite{goriely02}, the travelling waves that form in suspension bridges~\cite{chen97} and the rolling mode in vibrated thin plates~\cite{boudaoud07}.  

\begin{figure}
\centering
\includegraphics[width=7cm]{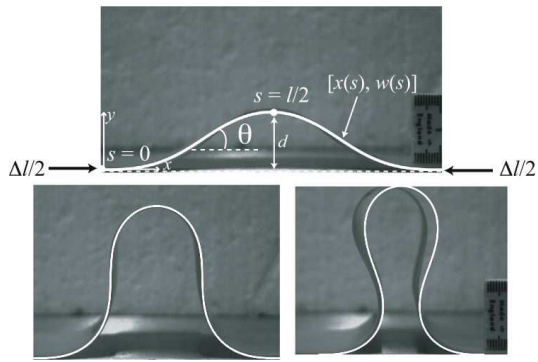}
\caption{The ruck shape observed in a sheet of natural rubber with thickness $h=0.75\mathrm{~mm}$ for different imposed end-end compressions $\dl$. Here the (non-dimensional) compressions are $\dL\equiv\dl/\lgr=0.3,2.7$ and $4.2$.}
\label{static_rucks}
\end{figure}

We begin by considering the static problem of a two-dimensional sheet of thickness $h$, density $\rho $ and bending stiffness $B$ lying on a rigid, horizontal substrate. A certain end-end displacement $\dl$ is then imposed symmetrically (see fig.~\ref{static_rucks}). Since compression of the sheet itself is energetically expensive, we expect that the sheet will buckle out of the plane. Opposing this is the weight per unit area of the sheet, $\rho gh$, which makes it energetically expensive for the sheet to lose contact with the substrate everywhere along its length (in contrast with the conventional elastica~\cite{love}). Instead, contact is lost only over a localized region --- a ruck is formed. We denote the shape of the ruck by $[x(s),w(s)]$ in which $s$ is the arc-length, though it is more convenient to determine an intrinsic equation for the local inclination of  the ruck, $\theta(s)$. This can be found by minimizing the value of the functional
\begin{equation}
{\cal F}\equiv \int_0^{l/2}\left\{\tfrac{1}{2}B\theta_s^2+\rho gh~w-\T\Bigl[\dl/l-\cos\theta\Bigr]\right\}~\upd s,
\label{eq:functional}
\end{equation} in which the first two terms are the bending and gravitational energies of deformation (per unit width), respectively, and the inextensibility constraint is enforced by a Lagrange multiplier $\T$ --- the compressive force per unit width in the ruck. Here $l$ is the total arc length contained in the ruck and we may consider only the interval $[0,l/2]$ by symmetry. Integrating by parts  and using $w_s=\sin\theta$, we find that $\int_0^{l/2}w~\upd s=\int_0^{l/2}\bigl(l/2-s\bigr)\sin\theta~\upd s$. We may then use the Euler-Lagrange equations to find a differential equation for the $\theta(s)$ that extremizes the functional in \eqref{eq:functional}. We first note that a natural lengthscale
\begin{equation}
\lgr\equiv\left(\frac{B}{\rho gh}\right)^{1/3}
\label{eq:ldefn}
\end{equation} arises. This elasto-gravitational length measures the typical sheet length required for the sheet to be deformed by its own weight. Introducing non-dimensional variables $S\equiv s/\lgr$, $\dL\equiv\dl/\lgr$, $L\equiv l/\lgr$ and $\ta \equiv \T\lgr^2/B$ the differential equation for $\theta(S)$ becomes
\begin{equation}
\theta_{SS}=-\ta \sin\theta+(L/2-S)\cos\theta,
\label{eq:heavyelast}
\end{equation} in which subscripts denote differentiation. Eqn.~\eqref{eq:heavyelast} is commonly referred to as the heavy elastica equation~\cite{wang86} --- it is the classical elastica equation~\cite{love} supplemented by a term representing the weight of the material. The differential equation \eqref{eq:heavyelast} is of second order but contains an unknown eigenvalue $\ta$ so that three boundary conditions are required. These are
\begin{equation}
\theta(0)=\theta_S(0)=0,\quad \theta(L/2)=0,
\label{eq:bcs}
\end{equation} which express that the sheet is horizontal to the substrate and torque free where it first touches the substrate (at $S=0$) as well as the symmetry of the sheet about $S=L/2$. The boundary value problem \eqref{eq:heavyelast}-\eqref{eq:bcs} may readily be solved numerically using, for example, MATLAB's \texttt{bvp4c} routine\cite{footnote1}. Some example profiles are shown in fig.~\ref{static_rucks} superimposed upon images of experimentally observed rucks in which a fixed end-end displacement $\dL$ is imposed. We observe very good agreement between the theoretically predicted and experimentally observed ruck shapes though for larger values of $\dL$ there is a symmetry breaking instability \cite{domokos03}. Although the ruck shapes shown in fig.~\ref{static_rucks} are qualitatively similar to those observed in the classical elastica~\cite{love}, a quantitative comparison shows that the heavy elastica shapes are flatter with a more localized `bump'. This is a result of the additional boundary condition $\theta_S(0)=0$ in the heavy elastica case.

Analytical progress may be made by linearizing  \eqref{eq:heavyelast} for small deformations, $\theta\ll1$. The resulting problem may be solved analytically giving
\begin{equation}
\ta \approx 4\alpha^2L^{-2},\quad L \approx\left(768/5\right)^{1/7}\alpha^{4/7}\dL^{1/7},
\label{eq:linres}
\end{equation} where $\alpha\approx 4.49341$ is the smallest positive solution of $\tan x=x$. Thus for a given displacement $\dL$ we have the laws $\ta\approx 3.441 \dL^{-2/7}$, $L\approx4.844\dL^{1/7}$ and $\delta\approx1.326\dL^{4/7}$ where $\delta\equiv d/\lgr$ is the dimensionless ruck height. These laws were previously presented as functions of $\ta$ \cite{wang81}. Recently, energy arguments for the scalings alone were also given \cite{kolinski09}. The dependence of $L$ on $\dL$ is distinct from the classical elastica where $L$ is the system size  and hence independent of $\dL$.

\begin{figure}
\centering
\psfrag{dl}[c]{$\dL\equiv \dl/\lgr$}
\psfrag{labelx}[c]{$\mu L_0$}
\includegraphics[width=6.5cm]{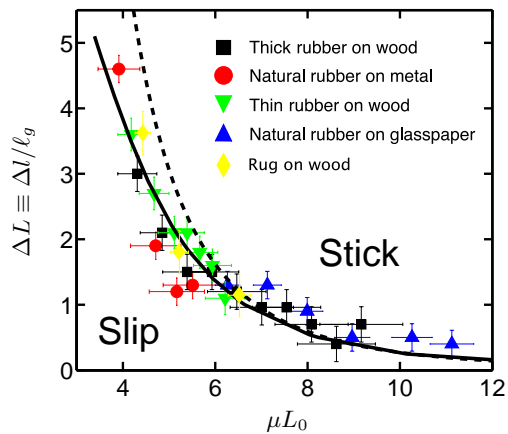}
\caption{(Color online) Regime diagram showing the regions of static friction--length $(\mu L_0)$, compression ($\dL$) parameter space for which a ruck is observed to stick (i.e.~remain) or slip (i.e.~flatten out) when the external compressive force is removed. The numerically computed boundary between sticking and slipping (solid curve) compares well with the asymptotic result \eqref{eq:lincrit} (dashed curve) for $\dL\ll1$. The unit of length is the elasto-gravitational length $\lgr$, defined in \eqref{eq:ldefn}.}
\label{regimed}
\end{figure}

We next consider the question will a ruck remain once the compressing force that formed it is removed? For a stationary ruck to stick, the compressive force within the ruck, $\ta $, must be smaller than the maximum static frictional force that can be exerted on the heavy sheet by the substrate. With a coefficient of static friction $\mu$, we therefore require that $\ta \leq\mu\bigl[(L_0-L)/2+N_0\bigr]$ in which $L_0$ is the total (non-dimensional) length of the sheet and $N_0$ is the normal reaction on the sheet at the point of contact between the sheet and substrate (i.e.~at $S=0$). Considering \eqref{eq:heavyelast} we see that $N_0=L/2$, i.e. the normal force at contact exactly balances the weight of the material contained in the half of the ruck for which $S<L/2$. We therefore find that a ruck sticks provided that $\ta \leq\mu L_0$. This result is exact though, in general, the dependence of $\ta $ on $\dL$ must be determined numerically. However, for small deformations, \eqref{eq:linres} may be used to relate $\ta $ and $\dL$ and show that a ruck sticks only if
\begin{equation}
\dL\geq \frac{20\sqrt{2}\alpha^3}{3}\bigl(\mu L_0\bigr)^{-7/2},
\label{eq:lincrit}
\end{equation}  i.e.~a ruck must be sufficiently large to stick. Qualitatively speaking, this result remains valid even for large deformations, since the compressive force $\ta $ is always a decreasing function of the imposed end-end compression, $\dL$. Fig.~\ref{regimed}  shows a regime diagram of the regions of $(\mu L_0,\dL)$ parameter space for which a ruck sticks or slips. Fig.~\ref{regimed} also shows the results of experiments in which the critical value of $\dL$ at which rucks first stick was determined for different sheet lengths using a variety of sheet materials and substrates. In each case, the value of $\lgr$ was measured experimentally using the loop test~\cite{stuart66} and $\mu$ was measured from the angle of friction. 

We now move on to discuss the dynamic motion of a ruck. The experiments described here involved mylar sheets (Goodfellow, Cambridge) of thicknesses $h=125,250,350$ and $500\mathrm{~\mu m}$ and length $2\mathrm{~m}$. These sheets were laid horizontally on various different substrates (to investigate the role of substrate friction). The far end of the sheet is left free to move\cite{footnote2} and the near end attached to a vertical track. We generate a moving ruck by lifting the near end of the sheet vertically along the track (allowing the free end to slide in) and then moving the near end rapidly downwards. This leads to the formation of a localized ruck, which typically propagates at speeds  $\sim1\mathrm{~m/s}$ away from the near end. The size of the ruck may be controlled by lifting the near end to different initial heights. However, to render air resistance negligible we ensure that the mass of air contained beneath the ruck is less than the mass of the ruck itself. This requires that $d\ll\rho h/\rho_a$ where $\rho_a$ is the density of air.

\begin{figure}
\centering
\psfrag{speed}{$c\sqrt{g\lgr}$}
\psfrag{labelx}[c]{$\lgr X_{peak}\mathrm{~(m)}$}
\psfrag{labelt}[c]{$(\lgr/g)^{1/2}T\mathrm{~(s)}$}
\psfrag{labelw}[c]{$W$}
\psfrag{newx}{$X-X_{peak}$}
\includegraphics[width=6.5cm]{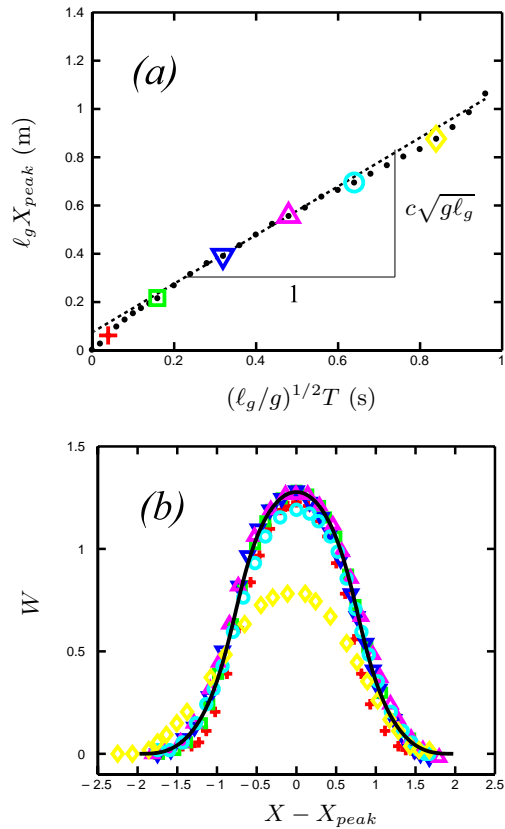}
\caption{(Color online) Experimental results for the dynamic propagation of a ruck in a mylar sheet ($h=125\mathrm{~\mu m}$). (a) The horizontal position of the peak of the ruck as a function of time (points). The speed in the steady state phase, $c\sqrt{g\lgr}$, is taken to be the gradient of the best fit line (dashed line) in the region where the shape is steady. (b) The shape of the travelling ruck at six instants of time. (The time of each profile is given by the position of the corresponding symbol in (a)). The solution of the heavy elastica equation \eqref{eq:heavyelast} with the same value of $\dL=1.05$ is shown by the solid curve.}
\label{dynexpts}
\end{figure}

Propagating rucks were filmed using a high speed camera (Photron Fastcam) with a frame rate of $250~\mathrm{Hz}$. (See EPAPS Document No. [number will be inserted by publisher] for a movie of a typical experiment.) The resulting images were analysed using ImageJ (NIH) to determine the ruck shape and the position of the peak of the ruck, $X_{peak}$, as functions of time. Typical experimental results are shown in fig.~\ref{dynexpts}. The time dependence of $X_{peak}$, fig.~\ref{dynexpts}a, shows that after some initial transient, the ruck moves at constant speed for a time before  it slows down and then speeds up. This latter phase of the motion corresponds to the free end of the sheet beginning to slip relative to the substrate. The ruck profiles shown in fig.~\ref{dynexpts}b show that while the ruck moves at constant speed its shape remains remarkably constant (though during the early transient and as the ruck  slips away its shape does evolve). We focus on understanding this phase of the motion here and leave the early and late time behaviors to a future investigation.

To derive the equations governing the dynamical motion of a ruck, we use the approach of \cite{coleman92,goriely02} albeit modified to include the vertical acceleration due to gravity. Resolving the stress within the sheet into horizontal, $f^{(h)}$, and vertical, $f^{(v)}$, components, we may write the horizontal and vertical force balances as
\begin{equation}
f^{(h)}_S=X_{TT},\quad f^{(v)}_S=Y_{TT}-1,
\label{eq:dynforce}
\end{equation} while the torque balance gives
\begin{equation}
\theta_{SS}-\frac{g\lgr}{E/\rho }\theta_{TT}=f^{(h)}Y_S-f^{(v)}X_S,
\label{eq:dyntheta}
\end{equation} in which $E$ is the Young's modulus of the sheet and $T\equiv t(g/\lgr)^{1/2}$ is dimensionless time.

The experimental results presented in fig.~\ref{dynexpts} suggest that there may be a travelling wave solution of eqns \eqref{eq:dynforce} and \eqref{eq:dyntheta}. It is therefore natural to transfer into a frame moving with constant speed $c$. We introduce a new variable $\eta\equiv S-cT$, which enables us to integrate \eqref{eq:dynforce} and determine the functions $f^{(h)}\equiv F^{(h)}(\eta)$ and $f^{(v)}\equiv F^{(v)}(\eta)$. Substituting these functions into \eqref{eq:dyntheta} gives a single equation for $\theta\equiv \Theta(\eta)$:
\begin{equation}
\left(1-c^2\rho g\lgr/E\right)\Theta_{\eta\eta}=-\bigl(\ta +c^2\bigr)\sin\Theta+(L/2-\eta)\cos\Theta.
\label{eq:wave}
\end{equation} In our experiments, the dimensional speed of the rucks, $c^2g\lgr\ll E/\rho $, the speed of sound within the sheet. We may therefore neglect the difference between the prefactor of $\Theta_{\eta\eta}$ in \eqref{eq:wave} and unity so that \eqref{eq:wave} becomes exactly the heavy elastica equation \eqref{eq:heavyelast} with the eigenvalue $\ta $ replaced by an `effective tension' $\ta +c^2$. Thus, for a given value of $\dL$, the shape of a steadily moving dynamic ruck must be exactly that of the static ruck with the same value of $\dL$. This point is illustrated in fig.~\ref{dynexpts}b where we see that while the ruck is propagating with a constant speed, its shape is indistinguishable from that of a static ruck with the same value of $\dL$.

It is a simple matter to solve numerically the eigenproblem \eqref{eq:wave} with boundary conditions analogous to \eqref{eq:bcs}. The linearized problem may be solved analytically giving the effective tension in terms of the ruck height $\delta\equiv d/\lgr$
\begin{equation}
\ta +c^2\approx3.962\delta^{-1/2}.
\label{eq:asyspeed}
\end{equation}

\begin{figure}
\centering
\psfrag{labely}[c]{$\ta+c^2$}
\psfrag{delta}[c]{$\delta$}
\psfrag{inset1}[c]{\small $d \mathrm{~(m)}$}
\psfrag{inset2}[c]{\small$\sqrt{g \lgr}c\mathrm{~(m/s)}$}
\includegraphics[width=6.5cm]{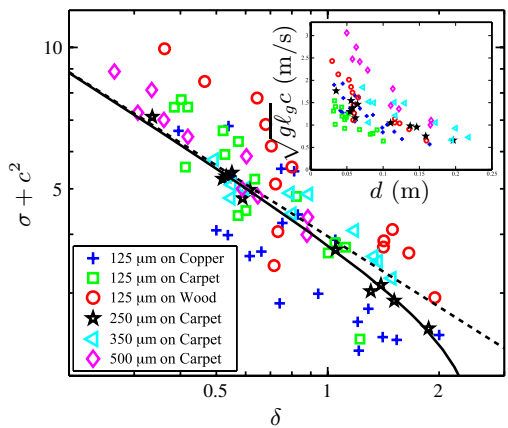}
\caption{(Color online) Main figure: Logarithmic plot of  the effective tension, $\ta +c^2$, (see text) measured experimentally for rucks in mylar sheets as a function of the  dimensionless ruck height, $\delta\equiv d/\lgr$. Experimental points are shown for different sheet thicknesses and substrates (hence different values of the friction coefficient $\mu$) as shown in the legend. The values measured of the friction coefficients are $\mu=0.13$ (copper), $\mu=0.41$ (carpet) and $\mu=0.16$ (wood). The dependence of $\ta +c^2$ on $\delta$ determined from the numerical solution of \eqref{eq:wave} is plotted as the solid curve while the asymptotic result \eqref{eq:asyspeed} is plotted as the dashed line. Inset: Linear plot of dimensional raw data showing the dependence of ruck speed $c\sqrt{g\lgr}$ on ruck height $d$.}
\label{modspeed}
\end{figure}

The presence of the combination $\ta +c^2$ in \eqref{eq:wave} and \eqref{eq:asyspeed} indicates that the ruck speed $c$ is not uniquely determined by the solution of the eigenproblem \eqref{eq:wave} with boundary conditions \eqref{eq:bcs}. This is demonstrated by the inset of fig.~\ref{modspeed}, which shows raw results for the speed, $c\sqrt{g\lgr}$, as a function of ruck height, $d$. However, we note that the slipping away of the ruck enables us to estimate the value of $\ta $  --- at the commencement of slipping,  $\ta $ must exactly balance the maximum static friction force that can be generated by the material remaining between the ruck and the free end. Fig.~\ref{modspeed} shows non-dimensional results for $\ta +c^2$ as a function of ruck height $\delta$ determined using this experimental procedure. Also plotted in fig.~\ref{modspeed} are the theoretical predictions obtained by solving the full problem numerically (solid curve) and the result of the linear analysis \eqref{eq:asyspeed} (dashed line). These show that we obtain good quantitative agreement between theory and experiment, though inaccuracies in determining the onset of sliding limit this agreement. 

In this Letter we have considered the properties of static rug rucks and shown that friction allows sufficiently large rucks to remain once the initial compression is removed. We have also considered the inertial dynamics of rucks, complementing a recent study of the creeping motion of a ruck on an inclined plane \cite{kolinski09}.  A result of particular interest is that large rucks generally move more slowly than smaller ones. If colliding rucks aggregate this fact would drive a population of travelling rucks to form one large, slow moving ruck. Future work will focus on determining whether this is qualitatively the same in populations of slip pulses in geophysical settings and could in turn lead to alternative mechanical rationalizations \cite{carlson94} of the statistics of earthquakes.

We thank L.~Mahadevan for bringing a related problem \cite{kolinski09} to our attention in 2003, arousing our interest in rucks. D.V.~is supported by the 1851 Royal Commission.


\begin{thebibliography}{20}
\bibitem{footnote1}{For simplicity, we take $L$ as given and calculate the corresponding $\dL$ using $\dL/2=\int_0^{L/2}\bigl(1-\cos\theta\bigr)~\upd S$.}
\bibitem{footnote2}{In some instances, a small weight ($\sim50\mathrm{~g}$) was attached to the free end to simulate an increased sheet length.}
\end{thebibliography}

\end{document}